\documentclass[12pt,a4paper]{article}

\usepackage{geometry}
\RequirePackage{amsmath}

\RequirePackage{mathrsfs}
\RequirePackage{amsfonts}
\RequirePackage{dsfont}
\RequirePackage{braket}
\RequirePackage{tabularx}
\usepackage{multirow}
\RequirePackage{nicefrac}
\RequirePackage{graphicx}
\RequirePackage{booktabs}
\RequirePackage{colortbl}
\RequirePackage{xcolor}
\RequirePackage[symbol]{footmisc}
\usepackage{mathtools}
\usepackage{comment}
\usepackage{blkarray}
\usepackage{stackengine}
\usepackage{float}
\usepackage{rotating}
\usepackage{hhline}
\usepackage{tikz}
\usepackage{multirow}
\usepackage[section]{placeins}
\newcommand{\boxalign}[2][0.97\textwidth]{
  \par\noindent\tikzstyle{mybox} = [draw=black,inner sep=6pt]
  \begin{center}\begin{tikzpicture}
   \node [mybox] (box){%
    \begin{minipage}{#1}{\vspace{-5mm}#2}\end{minipage}
   };
  \end{tikzpicture}\end{center}
}
\def\AEF{A.E. Faraggi}

\def\IJMP#1#2#3{{\it Int.\ J.\ Mod.\ Phys.}\/ {\bf A#1} (#2) #3}

\def\EJP#1#2#3{{\it Eur.\ Phys.\ Jour.}\/ {\bf C#1} (#2) #3}

\def\JHEP#1#2#3{{\it JHEP}\/ {\bf #1} (#2) #3}
\def\NPB#1#2#3{{\it Nucl.\ Phys.}\/ {\bf B#1} (#2) #3}
\def\PLB#1#2#3{{\it Phys.\ Lett.}\/ {\bf B#1} (#2) #3}
\def\PRD#1#2#3{{\it Phys.\ Rev.}\/ {\bf D#1} (#2) #3}
\def\PRL#1#2#3{{\it Phys.\ Rev.\ Lett.}\/ {\bf #1} (#2) #3}
\def\PRT#1#2#3{{\it Phys.\ Rep.}\/ {\bf#1} (#2) #3}
\def\etal{{\it et al\/}}

\def\beq{\begin{equation}}
\def\eeq{\end{equation}}
\def\beqn{\begin{eqnarray}}
\def\eeqn{\end{eqnarray}}
\def\ds{{$\tilde S$}}
\def\unahe{{${\overline{\rm NAHE}}$}}
\usepackage{empheq}
\usepackage[most]{tcolorbox}
\usepackage{blkarray}
\newtcbox{\mymath}[1][]{%
    nobeforeafter, math upper, tcbox raise base,
    enhanced, colframe=blue!30!black,
    colback=blue!30, boxrule=1pt,
    #1}

\newcommand{\CC}[2]{C{#1\atopwithdelims[]#2}}

\newcommand{\ba}{\begin{eqnarray}}
\newcommand{\ea}{\end{eqnarray}}
\DeclareRobustCommand{\sqbinom}{\genfrac[]{0pt}{}}

\numberwithin{equation}{section}

\begin{document}
\begin{titlepage}
\samepage{
\setcounter{page}{1}
\rightline{}
\rightline{November 2020}

\vfill
\begin{center}
  {\Large \bf{
      Type $\mathbf{\bar{0}}$ Heterotic String Orbifolds}}

\vspace{1cm}
\vfill

{\large Alon E. Faraggi\footnote{E-mail address: alon.faraggi@liv.ac.uk}, 
         Viktor G. Matyas\footnote{E-mail address: viktor.matyas@liv.ac.uk}
 and Benjamin Percival\footnote{E-mail address: benjamin.percival@liv.ac.uk} 
}
\\

\vspace{1cm}

{\it Dept.\ of Mathematical Sciences, University of Liverpool, Liverpool
L69 7ZL, UK\\}

\vspace{.025in}
\end{center}

\vfill
\begin{abstract}
\noindent

The $Z_2\times   Z_2$ heterotic string orbifold gives rise to
a large space of phenomenological three generation models that 
serves as a testing ground to explore how the Standard Model of particle
physics may be incorporated in a theory of quantum gravity.
Recently, we demonstrated the existence of type 0 $Z_2\times Z_2$
heterotic string orbifolds in which there are no massless fermionic
states. In this paper we demonstrate the existence of non--supersymmetric
tachyon--free $Z_2\times Z_2$ heterotic string orbifolds that do not contain any
massless bosonic states from the twisted sectors. 
We dub these configurations type ${\bar 0}$ models. They necessarily contain untwisted bosonic
states, producing the gravitational, gauge and scalar moduli degrees of
freedom, but possess an excess of massless fermionic states over
bosonic ones, hence producing a positive cosmological constant.
Such configurations may be instrumental when trying to understand
the string dynamics in the early universe.

\end{abstract}

\smallskip}

\end{titlepage}

\section{Introduction}\label{intro}

The Standard Model of particle physics provides viable parameterisation
of all sub--atomic observable phenomena. Yet, many enigmas remain, in
particular those pertaining to gravitational observations, {\it e.g.} the
dark matter and dark energy sectors. Furthermore, the general framework
that underlies the Standard Model, that of quantum field theories, is
fundamentally incompatible with general relativity,
the general framework that underlies gravitational observations.
String theory
is a contemporary framework that provides a perturbatively
consistent approach for the synthesis of quantum mechanics and
general relativity. In that context, it is sensible to construct
string models that aim to reproduce the general structure of the
Standard Model. In turn these string constructions may shed
light on some of the parameters of the Standard Model \cite{fhcp}.
For example, it has recently been proposed that modular symmetries
that are ubiquitous in string theory play a role in the flavour
structure of the Standard Model \cite{eclectic}. 

The $Z_2\times Z_2$ heterotic string orbifold gave rise to a large number
of phenomenological string models
\cite{fsu5,fny, alr,slm,lrs,acfkr,su62,frs,slmclass, lrsclass, lrsfertile}.
It has been studied primarily by using the
free fermionic formulation of the heterotic string in four
dimensions \cite{fff}, but bosonic constructions exist as well
\cite{stefanetal}. 
Detailed dictionaries between the two formalisms have been developed
\cite{z2xz2},
as well as tools to extract the smooth effective field  theory
limit of the orbifold constructions \cite{smooth}.
The $Z_2\times Z_2$ orbifold also exhibits rich mathematical
structures \cite{mathematical}, and it is of further interest
to explore how these are reflected in the phenomenological data
{\it e.g.} in the Standard Model flavour data.
The $Z_2\times Z_2$ orbifold compactifications
have been investigated in other string limits as well \cite{as}. 
The majority of phenomenological studies pertain to the
Standard Model particle data, but some cosmological scenarios
have been explored as well \cite{z2z2cosmo}. In this respect it
should be noted that contemporary understanding of string theory
is confined primarily to its static vacuum solutions and
dynamical questions are mostly explored in effective field
theory limits. This raises the prevailing enigma as to when, if at all, an effective field theory model can be incorporated into an
ultraviolet complete theory of quantum gravity. Or, more
concretely, when does an effective field theory
model have an embedding in string theory?

Developing a more complete understanding
of the string dynamics in the early universe requires elucidation of the 
non--supersymmetric as well as the tachyonic and unstable string configurations. 
%
Non--supersymmetric tachyon--free string vacua  
in ten dimensions were studied since the mid--eighties \cite{dh, kltclas, itoyama, gv}.
Some phenomenological studies in four dimensions of such vacua are explored in
\cite{nonsusy, interpol, aafs, ADM, FR}.
Recently, we embarked on the construction
of tachyon--free $Z_2\times Z_2$ heterotic string
models that are related to compactifications of the
ten dimensional non--supersymmetric and tachyonic string
vacua in ten dimensions \cite{spwsp,stable,so10tclass,PStclass}.
A tachyon--free three generation
Standard--like model in this class was presented in \cite{stable},
whereas in refs \cite{so10tclass} and \cite{PStclass} we performed
a broad classification of models with unbroken $SO(10)$ and
$SO(6)\times SO(4)$ unbroken GUT groups, respectively, as well as
the analysis of their vacuum energy, and models with equal numbers
of massless bosonic and fermionic degrees of freedom, {\it i.e.}
$N_b^0=N_f^0$. In ref. \cite{type0} the analysis of fermionic
$Z_2\times Z_2$ orbifolds was extended to type 0 models,
{\it i.e.} models that do not contain any massless fermionic
degrees of freedom. While clearly not of phenomenological relevance,
these cases are of particular interest in trying to develop a
picture of the string dynamics around unstable configurations,
due to their relative simplicity and high degree of symmetry. Indeed,
type 0 string constructions have been studied in other string theory
limits and their properties were explored \cite{type0string}.

In this paper we extend the analysis of such extreme
cases to tachyon--free heterotic string
models that do not contain any twisted bosonic
degrees of freedom. In analogy with type 0 models,
we refer to such configurations as type ${\bar 0}$ models.
It is apparent that such models contain untwisted bosonic degrees
of freedom that correspond to the gravitational, gauge and untwisted
scalar fields. However, in the type ${\bar 0}$
configurations that we present all
the bosonic degrees of freedom from the twisted sectors of the
$Z_2\times Z_2$ orbifold are projected out. As a consequence,
in such vacua there exist an excess of fermionic over bosonic
degrees of freedom and the models possess a positive cosmological constant.
Furthermore, in contrast to the type 0 models of ref. \cite{type0}
that necessarily contains some tachyonic degrees of freedom,
we find that most cases of type ${\bar0}$ models are free of tachyonic states.
We present type ${\bar 0}$ models that belong to the class of \ds--models
of ref. \cite{spwsp,stable,so10tclass, PStclass}, as well as the class of
$S$--models, where the first class are those models that descend from
a tachyonic ten dimensional vacuum, whereas the second are those that
can be regarded as compactifications of the non--supersymmetric $SO(16)\times
SO(16)$ ten dimensional tachyon--free vacuum. We also note the existence
of a supersymmetric vacuum that does not contain massless twisted bosons
that has indeed appeared in previous classifications \cite{fknr, fkr,
  acfkr, frs}. In these cases the partition function is vanishing,
whereas the type ${\bar 0}$ of interest are those that are
not supersymmetric, and with an excess of fermionic over
bosonic states. In such configurations the vacuum energy is positive.
Though they are unstable they may serve as laboratories to explore the possible
string dynamics in the early universe. We also remark that in all the type ${\bar 0}$
models that we find there are no spinorial or anti--spinorial representations of the $SO(10)$
GUT group, which is necessarily the case in the supersymmetric ${\bar 0}$ configurations. 

\section{Type $\mathbf{\bar{0}}$ $\mathbb{Z}_2\times \mathbb{Z}_2$
  Heterotic String Orbifold}\label{model}
We will utilize the free fermionic contruction \cite{fff} in which we require 
a set of boundary condition basis vectors and one--loop Generalised
GSO (GGSO) phases in order to define our models \cite{fff}.
The details of the formalism are not repeated here but we will be
adopting the conventional notation used in the
free fermionic constructions
\cite{fsu5, fny, alr, slm, lrs, acfkr, su62, frs, slmclass, lrsclass,
  lrsfertile}. 
  
The first type $\bar{0}$ model we found is built off the \unahe--set
that was employed in \cite{spwsp, stable}. In this set, the basis
vector $S$ that generates spacetime supersymmetry in NAHE--based models
\cite{nahe} is augmented with four periodic right--moving fermions,
which amounts to making the gravitinos massive. This introduces a general
$S\rightarrow {\tilde S}$ map in the space of models that was discussed in
detail in ref. \cite{stable, so10tclass, PStclass}.

The set of basis vectors is given by
\begin{align}\label{basis}
\mathds{1}&=\{\psi^\mu,\
\chi^{1,\dots,6},y^{1,\dots,6}, w^{1,\dots,6}\ | \ \overline{y}^{1,\dots,6},\overline{w}^{1,\dots,6},
\overline{\psi}^{1,\dots,5},\overline{\eta}^{1,2,3},\overline{\phi}^{1,\dots,8}\},\nonumber\\
\tilde{S}&=\{{\psi^\mu},\chi^{1,\dots,6} \ | \ \overline{\phi}^{3,4,5,6}\},\nonumber\\
{b_1}&=\{\psi^{\mu},\chi^{12},y^{34},y^{56}\; | \; \overline{y}^{34},
\overline{y}^{56},\overline{\eta}^1,\overline{\psi}^{1,\dots,5}\},\\
{b_2}&=\{\psi^{\mu},\chi^{34},y^{12},w^{56}\; | \; \overline{y}^{12},
\overline{w}^{56},\overline{\eta}^2,\overline{\psi}^{1,\dots,5}\},\nonumber\\
{b_3}&=\{\psi^{\mu},\chi^{56},w^{12},w^{34}\; | \; \overline{w}^{12},
\overline{w}^{34},\overline{\eta}^3,\overline{\psi}^{1,\dots,5}\},\nonumber\\
z_1&=\{\overline{\phi}^{1,\dots,4}\},\nonumber\\
G&=\{y^{1,...,6},w^{1,...,6} \ | \ \overline{y}^{1,...,6},\overline{w}^{1,...,6}\},\nonumber
\nonumber
\end{align}
and we further define the important linear combination
\beq 
z_2=1+b_1+b_2+b_3+z_1=\{\bar{\phi}^{5,6,7,8}\}.
\eeq
A model may then be specified through the assignment of modular invariant GGSO phases $\CC{v_i}{v_j}$ between the basis vectors. An example type $\bar{0}$ configuration arises for the GGSO assignment
{\begin{equation}
\small
\CC{v_i}{v_j}= 
\begin{blockarray}{ccccccccc}
&\mathbf{1}& \tilde{S} & b_1 & b_2&b_3&z_1 & G \\
\begin{block}{c(rrrrrrrr)}
\mathbf{1}& 1&   1&  -1&  -1&   1&   1&   1&\ \\
\tilde{S} & 1&  -1&   1&  -1&   1&  -1&  -1&\ \\
b_1       &-1&  -1&  -1&   1&   1&   1&  -1&\ \\
b_2       &-1&   1&   1&  -1&   1&   1&  -1&\ \\
b_3       & 1&  -1&   1&   1&   1&   1&   1&\ \\
z_1       &1&   1&   1&   1&   1&   1&  -1&\ \\ 
G         &1&  -1&  -1&  -1&   1&  -1&  -1&\ \\ 
\end{block}
\end{blockarray}
\label{ggsophases}
\end{equation}}

The model is free of (on--shell) tachyons and the gauge group is given by the model--independent contribution from the NS (untwisted) sector giving the vector bosons of $SO(10)\times U(1)^3\times SO(4)^3\times 
SU(2)^8$, as well as the additional gauge bosons arising from the presence of $\psi^\mu\ket{z_1+z_2}$ in the massless spectrum, as well as additional scalars from the $\{\lambda^a\}\{\bar{\lambda}^b\}\ket{z_k}$, $k=1,2$ and $\lambda^a$ is some left--moving oscillator not equal to $\psi^\mu$ and $\bar{\lambda}^b$ is any right--moving oscillator with NS boundary conditions in $z_k$. These additional scalars arise
in the untwisted sector necessarily to give the scalar moduli degrees of
freedom.
With the gauge enhancement the full gauge group of the model becomes 
\beq 
SO(10)\times U(1)^3\times SO(4)^3\times SO(8)^2. 
\eeq 
Apart from these untwisted
sector gauge bosons and scalars though, the massless spectrum contains exclusively fermionic states, 
as advertised for a type $\bar{0}$ configuration. These fermionic sectors are 
\begin{align}
    \begin{split}
        &\tilde{S}, \ \ 
        \tilde{S}+z_1, \ \  
        \tilde{S}+z_1+z_2, \ \  
        \tilde{S}+z_2, \\ 
        &b_1+b_2+b_3+G, \\  
        &\tilde{S}+b_i+b_j+G, \\ 
        &\tilde{S}+b_i+b_j+z_1+G, \\ 
        &\mathds{1}+\tilde{S}+b_i+G, \\ 
        &\mathds{1}+\tilde{S}+b_i+z_1+G, 
    \end{split}
\end{align}
where $i\neq j \neq k\in \{1,2,3\}$. This is notably all the possible fermionic massless sectors except $b_{1,2,3}$ which generate the $\mathbf{16}/\overline{\mathbf{16}}$ of $SO(10)$. 

Within the class of models with the minimal basis (\ref{basis}), 
possible twisted bosons may arise from the vectorial sectors
\begin{align}
    \begin{split}\label{VectBosons}
       V^1&=b_2+b_3+G,\\ 
       V^2&=b_1+b_3+G,\\ 
       V^3&=b_1+b_2+G, 
    \end{split}
\end{align}
which come with a right--moving oscillator, and the fermionic spinorial sectors
\begin{align}
    \begin{split}\label{FermBosons}
       B^1&=b_2+b_3+z_1+G,\\ 
       B^2&=b_1+b_3+z_1+G,\\ 
       B^3&=b_1+b_2+z_1+G, \\
       B^4&=\mathds{1}+b_1+z_1+G,\\ 
       B^5&=\mathds{1}+b_2+z_1+G,\\ 
       B^6&=\mathds{1}+b_3+z_1+G. 
    \end{split}
\end{align}
Type $\bar{0}$ models will be those in which the Hilbert space of GGSO-projected states $\ket{S_\xi}$ 
\begin{equation}
    \mathcal{H}=\bigoplus_{\xi\in\Xi}\prod^{k}_{i=1}
\left\{ e^{i\pi v_i\cdot F_{\xi}}\ket{S_\xi}=\delta_{\xi}
\CC{\xi}{v_i}^*\ket{S_\xi}\right\}, 
\end{equation}
only has contributions from sectors $\xi$ in the additive space $\Xi$ with fermionic spin statistic index $\delta_{\xi}=-1$ at the massless level, except for the aforementioned untwisted sectors. Thus, using GGSO projections, we can derive the conditions on the GGSO phases in order to realise type $\bar{0}$ configurations. 

One easy way to derive these conditions is to first inspect the projection of the sector $B^4$ which can only be projected by $z_1$ such that
\beq \label{B4cond1}
\CC{\mathds{1}+b_1+z_1+G}{z_1}=-1 \ \ \iff \ \ \CC{z_1}{b_1}\CC{z_1}{G}=-1. 
\eeq 
Similarly, projecting $B^5$ and $B^6$ requires
\beq \label{B4cond2}
\CC{z_1}{b_2}\CC{z_1}{G}=-1=\CC{z_1}{b_3}\CC{z_1}{G}.
\eeq 
The projection of $B^1$ then requires that
\beq 
\left( 1+\CC{B^1}{b_1+G}\right)\left( 1+\CC{B^1}{z_2}\right)=0,
\eeq 
 and so expanding $z_2$ in terms of basis vectors and using the ABK rules for these two phases results in
\begin{align}
\begin{split}\label{B1ABK}
        \CC{B^1}{b_1+G}&=-\CC{b_2}{b_1}\CC{b_2}{G}\CC{b_3}{b_1}\CC{b_3}{G}\underbrace{\CC{z_1}{b_1}\CC{z_1}{G}}_{\text{=$-1$ from (\ref{B4cond1})}}\CC{G}{b_1}\CC{G}{G} \\
        &=-\CC{b_2}{b_1}\CC{b_3}{b_1}\CC{1}{G}\CC{b_1}{G}\CC{b_2}{G}\CC{b_3}{G} \\
        \CC{B^1}{z_2}&=\CC{z_2}{b_2}\CC{z_2}{b_3}\CC{z_2}{z_1}\CC{z_2}{G}\\
        &=-\CC{b_2}{b_1}\CC{b_3}{b_1}\CC{1}{G}\CC{b_1}{G}\CC{b_2}{G}\CC{b_3}{G},    
\end{split}
\end{align}
\textit{i.e.} they are equal. Considering also projecting $B^2$ and $B^3$ we can therefore deduce 
\beq \label{B1cond1}
\CC{b_1}{b_2}=\CC{b_1}{b_3}=\CC{b_2}{b_3}
\eeq 
and
\beq \label{B1cond2}
\CC{1}{G}\CC{b_1}{G}\CC{b_2}{G}\CC{b_3}{G}=1.
\eeq 
Finally, we can note that the GGSO phases that can project on the $V^1$ sector are
\beq 
O_{V^1}=\left\{\CC{V^1}{z_1},\CC{V^1}{z_2},\CC{V^1}{b_1+G}\right\},
\eeq 
and since $B^1=V^1+z_1$ these can be simplified using (\ref{B1ABK}) and (\ref{B4cond1}) to give
\beq 
O_{V^1}=\left\{\CC{z_1}{G},\CC{z_1}{G},-1\right\}.
\eeq 
We can write the projection condition for all possible oscillators as 
\beq 
\# \left\{x \in O_{V^1} | x=-1\right\} \neq 1,
\eeq 
therefore we observe that $\CC{z_1}{G}=-1$ for the projection of $V^1$. Using this in equations (\ref{B4cond1}) and (\ref{B4cond2}) and rewriting conditions (\ref{B1cond1}) and (\ref{B1cond2}) we get the full conditions for the type $\bar{0}$ string vacua
\boxalign{\begin{align}
\label{Type0barConds}
 \CC{z_1}{b_1}&=\CC{z_1}{b_2}=\CC{z_1}{b_3}=1, \ \ \ \ \CC{z_1}{G}=-1,  \\  
 \CC{b_1}{b_2}&=\CC{b_1}{b_3}, \ \ \ \  \CC{b_2}{b_3}=\CC{b_1}{b_3}, \\ 
 \CC{1}{G}&= \CC{b_1}{G}\CC{b_2}{G}\CC{b_3}{G}.
\end{align}}
Therefore we see that 7 GGSO phases are fixed and we have 14 free phases. Similar constraints were derived for type 0 models in ref. \cite{type0} where it was shown that in a similar minimal basis to (\ref{basis}) there were 12 free phases giving $2^{12}=4096$ versions of a single unique type 0 
partition function. 
To check whether we have $2^{14}$ versions of a unique 
partition function or not in our type $\bar{0}$ case we must analyse the partition function, which in free fermionic models is given by the integral
\begin{equation}
    Z = \int_\mathcal{F}\frac{d^2\tau}{\tau_2^2}\, Z_B  \sum_{\alpha,\beta} \CC{\alpha}{\beta} \prod_{f} Z \sqbinom{\alpha(f)}{\beta(f)}=  \int_\mathcal{F} \frac{d^2\tau}{\tau_2^3} \,\sum_{n.m} a_{mn}\, q^m \bar{q}^n,
    \label{ZInt}
\end{equation}
where $d^2\tau/\tau_2^2$ is the modular invariant measure and $Z_B$ denotes the contribution from the worldsheet bosons. The product is over the free worldsheet fermions.  On the right hand side of (\ref{ZInt}) we have expanded the partition function in terms of the parameters $q\equiv e^{2\pi i \tau}$ and $\bar{q}\equiv e^{-2\pi i \bar{\tau}}$, which allows us to read off the boson-fermion degeneracies at each mass level. That is, $a_{mn}=N_b-N_f$ at mass level $(m,n)$ and so we expect that type $\bar{0}$ models have large negative $a_{00}$ due to the absence of twisted bosonic states. Throughout this paper we will refer to the unintegrated sum as the partition function. The whole integrated expression (\ref{ZInt}) represents the one-loop worldsheet vacuum energy $\Lambda_{\text{WS}}$ of our theory and thus is a dimensionless quantity. It is related to the 4D spacetime cosmological constant $\Lambda$ via 
\beq\label{lambda}
\Lambda =  -\frac{1}{2}\mathcal{M}^4\Lambda_{\text{WS}},
\eeq
where $\mathcal{M}$ is given in terms of the string mass as $\mathcal{M}=M_{String}/2\pi$. In the following, when we refer to the cosmological constant, we implicitly mean the spacetime value, but for simplicity we drop the factor of $\mathcal{M}^4/2$. This can be reinstated if needed based on dimensional analysis.

Performing the calculation of the partition function for the $2^{14}=16384$ type $\bar{0}$ configurations we find that they all share the partition function
\beq 
Z=2\,q^0\bar{q}^{-1}-728\,q^0\bar{q}^0+288\,q^{1/2}\bar{q}^{-1/2}+1088\,q^{-1/2}\bar{q}^{1/2}+38400\,q^{1/2}\bar{q}^{1/2}+\cdots,
\eeq 
and so are, indeed, the same model. We note that there are no on-shell tachyons and the absence of twisted bosons ensures a large negative contribution at the massless level $N_b^0-N_f^0=-728$. We can calculate the cosmological constant now for this unique case. Due to the abundance of fermionic states compared to bosonic ones, we expect a positive cosmological constant, and performing the modular integral using standard techniques we, indeed, find
\beq 
\Lambda=238.38\times\mathcal{M}^4.
\eeq 

In ref. \cite{type0} it was shown that type 0 models exhibit misaligned supersymmetry \cite{MSUSY}, and further details of this behaviour were given. Similarly, all type $\bar{0}$ models presented in this paper exhibit a form of misaligned supersymmetry, meaning that the boson-fermion degeneracies oscillate while ascending through the KK tower of massive states. 

\section{Generalised Type $\bar{0}$ $\tilde{S}$-models}
In order to do a more general search for type $\bar{0}$
models we can generalise from the basis (\ref{basis}) to
\begin{align}\label{basisStTi}
\mathds{1}&=\{\psi^\mu,\
\chi^{1,\dots,6},y^{1,\dots,6}, w^{1,\dots,6}\ | \ \overline{y}^{1,\dots,6},\overline{w}^{1,\dots,6},
\overline{\psi}^{1,\dots,5},\overline{\eta}^{1,2,3},\overline{\phi}^{1,\dots,8}\},\nonumber\\
\tilde{S}&=\{{\psi^\mu},\chi^{1,\dots,6} \ | \ \overline{\phi}^{3,4,5,6}\},\nonumber\\
{T_1}&=\{y^{1,2},w^{1,2}\; | \; \overline{y}^{1,2},\overline{w}^{1,2}\},\nonumber\\ 
{T_2}&=\{y^{3,4},w^{3,4}\; | \; \overline{y}^{3,4},\overline{w}^{3,4}\},\nonumber\\ 
{T_3}&=\{y^{5,6},w^{5,6}\; | \; \overline{y}^{5,6},\overline{w}^{5,6}\},\nonumber\\ 
{b_1}&=\{\psi^{\mu},\chi^{12},y^{34},y^{56}\; | \; \overline{y}^{34},
\overline{y}^{56},\overline{\eta}^1,\overline{\psi}^{1,\dots,5}\},\\
{b_2}&=\{\psi^{\mu},\chi^{34},y^{12},w^{56}\; | \; \overline{y}^{12},
\overline{w}^{56},\overline{\eta}^2,\overline{\psi}^{1,\dots,5}\},\nonumber\\
{b_3}&=\{\psi^{\mu},\chi^{56},w^{12},w^{34}\; | \; \overline{w}^{12},
\overline{w}^{34},\overline{\eta}^3,\overline{\psi}^{1,\dots,5}\},\nonumber\\
z_1&=\{\overline{\phi}^{1,\dots,4}\},\nonumber
\nonumber
\end{align}
where introducing $T_i$, $i=1,2,3$ allows for internal symmetric shifts around the 3 internal $T^2$ tori. Since we have 9 basis vectors there are $2^{9(9-1)/2}=2^{36}\sim 6.87 \times 10^{10}$ independent GGSO phase configurations.

The bosonic sectors that need projecting in this basis are similar to (\ref{VectBosons}), up to allowing for the shifts induced by $T_i$. Explicitly, there are 15 vectorial bosonic sectors
\begin{align}
    \begin{split}\label{VectBosonsTi}
       V^1_{pq}&=b_2+b_3+T_1+pT_2+qT_3,\\ 
       V^2_{pq}&=b_1+b_3+T_2+pT_1+qT_3,\\ 
       V^3_{pq}&=b_1+b_2+T_3+pT_1+qT_2,\\
       V^4&=T_1+T_2, \\
       V^5&=T_1+T_3, \\
       V^6&=T_2+T_3, 
    \end{split}
\end{align}
which come with a right--moving oscillator and $p,q=0,1$. Additionally, there are 30 fermionic spinorial sectors
\begin{align}
    \begin{split}\label{FermBosonsTi}
       B^1_{pq}&=b_2+b_3+z_1+T_1+pT_2+qT_3,\\ 
       B^2_{pq}&=b_1+b_3+z_1+T_2+pT_1+qT_3,\\ 
       B^3_{pq}&=b_1+b_2+z_1+T_3+pT_1+qT_2, \\
       B^4_{pq}&=\mathds{1}+b_1+z_1+T_1+pT_2+qT_3,\\ 
       B^5_{pq}&=\mathds{1}+b_2+z_1+T_2+pT_1+qT_3,\\ 
       B^6_{pq}&=\mathds{1}+b_3+z_1+T_3+pT_1+qT_2,\\ 
       B^7&=T_1+T_2+z_1,\\
       B^8&=T_1+T_3+z_1, \\
       B^9&=T_2+T_3+z_1,\\ 
       B^{10}&=T_1+T_2+z_2,\\
       B^{11}&=T_1+T_3+z_2, \\
       B^{12}&=T_2+T_3+z_2.\\ 
    \end{split}
\end{align}
Implementing the GGSO projection conditions on all the sectors and scanning over $10^8$ random GGSO phase configurations resulted in uncovering 5676 type $\bar{0}$ configurations that correspond to just two distinct tachyon--free models and two distinct tachyonic models. The first tachyon--free model has partition function
\beq 
Z=2\,q^0\bar{q}^{-1}-440\,q^0\bar{q}^0+32\,q^{1/4}\bar{q}^{-3/4}-6080\,q^{1/4}\bar{q}^{1/4}+\cdots,
\eeq 
and cosmological constant
\beq 
\Lambda =213.27\times\mathcal{M}^4.
\eeq 
Whereas the second tachyon--free model has partition function 
\beq 
Z=2\,q^0\bar{q}^{-1}-504\,q^0\bar{q}^0+48\,q^{1/4}\bar{q}^{-3/4}-12192\,q^{1/4}\bar{q}^{1/4}+\cdots,
\eeq 
and cosmological constant
\beq 
\Lambda =278.60\times\mathcal{M}^4.
\eeq 
Both models contain the same gauge boson enhancement and additional scalars from the sectors $z_1,z_2$ and $z_1+z_2$ as in case with minimal basis (\ref{basis}). Other than these untwisted bosons the two models contain only twisted fermionic states in their massless spectra, as required for type $\bar{0}$ configurations. 

Regarding the two tachyonic models, we have one model with partition function
\beq 
Z=2\,q^0\bar{q}^{-1}+32q^{-1/4}\bar{q}^{-1/4}-1016\,q^0\bar{q}^0+4096\,q^{1/4}\bar{q}^{1/4}+\cdots,
\eeq 
which has 32 tachyonic states and one with partition function
\beq 
Z=2\,q^0\bar{q}^{-1}+48\,q^{-1/4}\bar{q}^{-1/4}-1272\,q^0\bar{q}^0+5120\,q^{1/4}\bar{q}^{1/4}+\cdots,
\eeq 
which has 48 tachyonic states. Such models with a tachyonic instability should not be written of as of no interest. In particular, moving away from the free fermionic point in the moduli space or considering the theory in a different background  may stabilise the model. Furthermore, there may be ways to connect such unstable vacua to stable ones via interpolation. 
\section{Generalised Type $\bar{0}$ $S$-models}
We can now do a similar exploration of type $\bar{0}$ models within a class of models which include the SUSY generating basis vector $S$. We employ a very familiar choice of $SO(10)$ basis 
\begin{align}\label{basisSTi}
\mathds{1}&=\{\psi^\mu,\
\chi^{1,\dots,6},y^{1,\dots,6}, w^{1,\dots,6}\ | \ \overline{y}^{1,\dots,6},\overline{w}^{1,\dots,6},
\overline{\psi}^{1,\dots,5},\overline{\eta}^{1,2,3},\overline{\phi}^{1,\dots,8}\},\nonumber\\
S&=\{{\psi^\mu},\chi^{1,\dots,6} \},\nonumber\\
{T_1}&=\{y^{1,2},w^{1,2}\; | \; \overline{y}^{1,2},\overline{w}^{1,2}\},\nonumber\\ 
{T_2}&=\{y^{3,4},w^{3,4}\; | \; \overline{y}^{3,4},\overline{w}^{3,4}\},\nonumber\\ 
{T_3}&=\{y^{5,6},w^{5,6}\; | \; \overline{y}^{5,6},\overline{w}^{5,6}\},\nonumber\\ 
{b_1}&=\{\chi^{3,4,5,6},y^{34},y^{56}\; | \; \overline{y}^{34},
\overline{y}^{56},\overline{\eta}^1,\overline{\psi}^{1,\dots,5}\},\\
{b_2}&=\{\chi^{1,2,5,6},y^{12},y^{56}\; | \; \overline{y}^{12},
\overline{y}^{56},\overline{\eta}^2,\overline{\psi}^{1,\dots,5}\},\nonumber\\
z_1&=\{\overline{\phi}^{1,\dots,4}\},\nonumber\\
z_2&=\{\overline{\phi}^{5,\dots,8}\},\nonumber
\nonumber
\end{align}
which is exactly the same as that used to classify non--SUSY string models in ref. \cite{FR}. We will note the important linear combination in this basis
\beq 
x=1+S+\sum_{i=1,2,3}T_i + \sum_{k=1,2}z_k,
\eeq 
and then have the combination $b_3=b_1+b_2+x$. As in the $\tilde{S}$-models we have 9 basis vectors and so the number of independent GGSO phase configurations is $2^{9(9-1)/2}=2^{36}\sim 6.87 \times 10^{10}$.

A key difference between this basis and the basis (\ref{basisStTi}) is that there exists a supersymmetric subspace of the full space for certain choices of GGSO phase. In particular, the $S$ sector generates supersymmetry whenever
\beq \label{susyPhases}
\CC{S}{T_i}=\CC{S}{z_k}=-1, \ \ \ \ \ \ \ i=1,2,3 \text{ and } k=1,2
\eeq 
which, furthermore, automatically ensures the projection of tachyonic sectors through the $S$ GGSO projection. 

Now we turn to the massless bosonic vectorial sectors that in our $S$--models arise from
\beqn
\label{SVectFerms}
& & b_i+x+pT_j+qT_k,\nonumber\\
& &T_1+T_2,\\
& &T_1+T_3, \nonumber\\
& &T_2+T_3,\nonumber
\eeqn
and the massless bosonic spinorial sectors from
\beqn
\label{SSpinFerms}
& &b_i+pT_j+qT_k,\\
& &b_i+x+z_1+pT_j+qT_k, \nonumber\\
& &b_i+x+z_2+pT_j+qT_k,\nonumber\\
& &T_1+T_2+z_1,\nonumber\\
& &T_1+T_3+z_1, \\
& &T_2+T_3+z_1,\nonumber\\
& &T_1+T_2+z_2,\nonumber\\
& &T_1+T_3+z_2, \\
& &T_2+T_3+z_2,\nonumber
\eeqn
where $i\neq j \neq k\in \{1,2,3\}$ and $p,q\in \{0,1\}$. 

We can now search for type $\bar{0}$ GGSO configurations by implementing the conditions for the GGSO projection of all these massless twisted bosonic sectors.

In a random scan of $10^8$ independent GGSO phase configurations we found one supersymmetric model which contains a very simple massless spectrum containing the untwisted gauge bosons from the NS sector and its gauginos from the $S$ sector, along with gauge enhancements and additional scalars of some form from $z_1,z_2,z_1+z_2$ and $x$ and their superpartners from  $S+z_1,S+z_2,S+z_1+z_2$ and $S+x$, respectively. The other type $\bar{0}$ models arising in our $10^8$ scan are non--supersymmetric. 

All the type $\bar{0}$ models are summarised in Table \ref{STable} with their partition functions, key characteristics and frequency within the sample delineated. Where we recall that the frequency refers to the number of different GGSO phase configurations corresponding to the same partition function. The projected total number is simply how many we expect in the full space of $2^{36}$ independent GGSO phase configurations. In principle, the exact constraints on the GGSO phases for each model could be deriving and the free phases found to derive the exact number of each model in the total space.
\begin{table}[!ht]
\centering
\small
\setlength{\tabcolsep}{5pt}
\def\arraystretch{1.2}
\begin{tabular}{|c|c|c|c|c|c|}
\hline 
\multirow{2}{*}{\textbf{Partition Function }}& \multirow{2}{*}{$\Lambda \, [\mathcal{M}^4]$}&\multirow{2}{*}{\textbf{Tachyons?}}&\multirow{2}{*}{\textbf{SUSY?}}&\textbf{\# Models }&\textbf{Total  \#}\\
&&&&\textbf{in Sample}&\textbf{Projected}\\
\hline
\multirow{2}{*}{$Z=0$}&\multirow{2}{*}{0}&\multirow{2}{*}{No}&\multirow{2}{*}{Yes}&\multirow{2}{*}{562}&\multirow{2}{*}{$3.86\times 10^5$}\\
&&&&&\\
\hline
$Z=2\bar{q}^{-1}-632+48q^{1/4}\bar{q}^{-3/4}$&\multirow{2}{*}{293.8}&\multirow{2}{*}{No}&\multirow{2}{*}{No}&\multirow{2}{*}{389}&\multirow{2}{*}{$2.67\times 10^5$}\\
$-8096q^{1/4}\bar{q}^{1/4}+\cdots$&&&&&\\
\hline
$Z=2\bar{q}^{-1}-120+32q^{1/4}\bar{q}^{-3/4}$&\multirow{2}{*}{125.6}&\multirow{2}{*}{No}&\multirow{2}{*}{No}&\multirow{2}{*}{284}&\multirow{2}{*}{$1.95\times 10^5$}\\
$-6080q^{1/4}\bar{q}^{1/4}+\cdots$&&&&&\\
\hline
$Z=2\bar{q}^{-1}-568+32q^{1/4}\bar{q}^{-3/4}$&\multirow{2}{*}{223.97}&\multirow{2}{*}{No}&\multirow{2}{*}{No}&\multirow{2}{*}{1163}&\multirow{2}{*}{$7.99\times 10^5$}\\
$-1984q^{1/4}\bar{q}^{1/4}+\cdots$&&&&&\\
\hline
$Z=2\bar{q}^{-1}-504+32q^{1/4}\bar{q}^{-3/4}$&\multirow{2}{*}{158.64}&\multirow{2}{*}{No}&\multirow{2}{*}{No}&\multirow{2}{*}{715}&\multirow{2}{*}{$3.91\times 10^5$}\\ 
$+4128q^{1/4}\bar{q}^{1/4}+\cdots$&&&&&\\
\hline
$Z=2\bar{q}^{-1}+32q^{-1/4}\bar{q}^{-1/4}-664$&\multirow{2}{*}{$\infty$}&\multirow{2}{*}{Yes}&\multirow{2}{*}{No}&\multirow{2}{*}{287}&\multirow{2}{*}{$1.97\times 10^5$}\\
$+6144q^{1/4}\bar{q}^{1/4}+\cdots$&&&&&\\
\hline
$Z=2\bar{q}^{-1}+32q^{-1/4}\bar{q}^{-1/4}-1272$&\multirow{2}{*}{$\infty$}&\multirow{2}{*}{Yes}&\multirow{2}{*}{No}&\multirow{2}{*}{290}&\multirow{2}{*}{$1.99\times 10^5$}\\
$+5888^{1/4}\bar{q}^{1/4}+\cdots$&&&&&\\
\hline
$Z=2\bar{q}^{-1}+32q^{-1/4}\bar{q}^{-1/4}-632$&\multirow{2}{*}{$\infty$}&\multirow{2}{*}{Yes}&\multirow{2}{*}{No}&\multirow{2}{*}{301}&\multirow{2}{*}{$2.07\times 10^5$}\\
$-512q^{1/4}\bar{q}^{1/4}+\cdots$&&&&&\\
\hline
$Z=2\bar{q}^{-1}+32q^{-1/4}\bar{q}^{-1/4}-1528$&\multirow{2}{*}{$\infty$}&\multirow{2}{*}{Yes}&\multirow{2}{*}{No}&\multirow{2}{*}{429}&\multirow{2}{*}{$2.95\times 10^5$}\\
$+4608q^{1/4}\bar{q}^{1/4}+\cdots$&&&&&\\
\hline
$Z=2\bar{q}^{-1}+32q^{-1/4}\bar{q}^{-1/4}-1528$&\multirow{2}{*}{$\infty$}&\multirow{2}{*}{Yes}&\multirow{2}{*}{No}&\multirow{2}{*}{395}&\multirow{2}{*}{$2.71\times 10^5$}\\
$+11008q^{1/4}\bar{q}^{1/4}+\cdots$&&&&&\\
\hline
$Z=2\bar{q}^{-1}+48q^{-1/4}\bar{q}^{-1/4}-1016$&\multirow{2}{*}{$\infty$}&\multirow{2}{*}{Yes}&\multirow{2}{*}{No}&\multirow{2}{*}{155}&\multirow{2}{*}{$1.07\times 10^5$}\\
$-1792q^{1/4}\bar{q}^{1/4}+\cdots$&&&&&\\
\hline
$Z=2\bar{q}^{-1}+144q^{-1/4}\bar{q}^{-1/4}-504$&\multirow{2}{*}{$\infty$}&\multirow{2}{*}{Yes}&\multirow{2}{*}{No}&\multirow{2}{*}{153}&\multirow{2}{*}{$1.05\times 10^5$}\\
$+9472^{1/4}\bar{q}^{1/4}+\cdots$&&&&&\\
\hline
\end{tabular}
\caption{\label{Stype0bar}\emph{Summary of type $\bar{0}$ models arising from the basis (\ref{basisSTi}). The cosmological constant $\Lambda$ is expressed in units of $\mathcal{M}^4$ as in (\ref{lambda}).}}\label{STable}
\end{table}

\section{Conclusions}

In this paper we explored the existence of $Z_2\times Z_2$ heterotic string
orbifolds that do not contain any massless spacetime scalar bosons from
the twisted sectors. In analogy with the type 0 $Z_2\times Z_2$
heterotic string orbifolds that were presented in \cite{type0},
we dubbed such configurations type ${\bar 0}$ models. We presented
two classes of such models, where the first are of the \ds--models type,
whereas the second belong to the class of $S$--models. We note that the
second class also contains a supersymmetric model that necessarily
do not contain twisted fermionic states and have vanishing
cosmological constant, whereas all other type $\bar{0}$ models found in 
both classes are non--supersymmetric and necessarily have an excess of
fermionic over bosonic states and therefore have a positive
cosmological constant. While our findings at this stage should be
regarded as mere curiosities, it is plausible that they may
contribute to the understanding of the string dynamics in the
early universe. We have also found that in all the type ${\bar 0}$
models, there are no spinorial or anti--spinorial representations of the $SO(10)$
GUT group. This is necessarily the case in the supersymmetric ${\bar 0}$ configurations, 
which therefore necessarily have a vanishing Euler characteristic. 
The non--supersymmetric ${\bar 0}$ configurations may therefore be interpreted as 
supersymmetric ${\bar 0}$ models, in which supersymmetry is maximally violated. 
A feature that may be explored by studying the interpolations between the two cases.

\section*{Acknowledgments}

The work of VGM is supported in part by EPSRC grant EP/R513271/1.
The work of BP is supported in part by STFC grant ST/N504130/1.

\bigskip

\bibliographystyle{unsrt}

\end{document}